\documentstyle[12pt]{article}

\def\hybrid{\topmargin -20pt	\oddsidemargin 0pt
	\headheight 0pt	\headsep 0pt
	\textwidth 6.45in	
	\textheight 9.5in	
	\marginparwidth .875in
	\parskip 5pt plus 1pt	\jot = 1.5ex}

\hybrid

\catcode`\@=11

\def\marginnote#1{}
\newcount\hour
\newcount\minute
\newtoks\amorpm
\hour=\time\divide\hour by60
\minute=\time{\multiply\hour by60 \global\advance\minute by-\hour}
\edef\standardtime{{\ifnum\hour<12 \global\amorpm={am}%
	\else\global\amorpm={pm}\advance\hour by-12 \fi
	\ifnum\hour=0 \hour=12 \fi
	\number\hour:\ifnum\minute<10 0\fi\number\minute\the\amorpm}}
\edef\militarytime{\number\hour:\ifnum\minute<10 0\fi\number\minute}

\def\draftlabel#1{{\@bsphack\if@filesw {\let\thepage\relax
   \xdef\@gtempa{\write\@auxout{\string
      \newlabel{#1}{{\@currentlabel}{\thepage}}}}}\@gtempa
   \if@nobreak \ifvmode\nobreak\fi\fi\fi\@esphack}
	\gdef\@eqnlabel{#1}}
\def\@eqnlabel{}
\def\@vacuum{}
\def\draftmarginnote#1{\marginpar{\raggedright\scriptsize\tt#1}}

\def\draft{\oddsidemargin -.5truein
	\def\@oddfoot{\sl preliminary draft \hfil
	\rm\thepage\hfil\sl\today\quad\militarytime}
	\let\@evenfoot\@oddfoot	\overfullrule 3pt
	\let\label=\draftlabel
	\let\marginnote=\draftmarginnote
   \def\@eqnnum{(\theequation)\rlap{\kern\marginparsep\tt\@eqnlabel}%
\global\let\@eqnlabel\@vacuum}  }

\def\preprint{\twocolumn\sloppy\flushbottom\parindent 2em
	\leftmargini 2em\leftmarginv .5em\leftmarginvi .5em
	\oddsidemargin -.5in	\evensidemargin -.5in
	\columnsep .4in	\footheight 0pt
	\textwidth 10.in	\topmargin  -.4in
	\headheight 12pt \topskip .4in
	\textheight 6.9in \footskip 0pt
	\def\@oddhead{\thepage\hfil\addtocounter{page}{1}\thepage}
	\let\@evenhead\@oddhead	\def\@oddfoot{}	\def\@evenfoot{} }

\def\numberbysection{\@addtoreset{equation}{section}
	\def\theequation{\thesection.\arabic{equation}}}

\def\underline#1{\relax\ifmmode\@@underline#1\else
	$\@@underline{\hbox{#1}}$\relax\fi}

\def\titlepage{\@restonecolfalse\if@twocolumn\@restonecoltrue\onecolumn
     \else \newpage \fi \thispagestyle{empty}\c@page\z@
	\def\thefootnote{\fnsymbol{footnote}} }

\def\endtitlepage{\if@restonecol\twocolumn \else \newpage \fi
	\def\thefootnote{\arabic{footnote}}
	\setcounter{footnote}{0}}  

\catcode`@=12
\relax

\def\figcap{\section*{Figure Captions\markboth
	{FIGURECAPTIONS}{FIGURECAPTIONS}}\list
	{Figure \arabic{enumi}:\hfill}{\settowidth\labelwidth{Figure 999:}
	\leftmargin\labelwidth
	\advance\leftmargin\labelsep\usecounter{enumi}}}
 \relax
\def\tablecap{\section*{Table Captions\markboth
	{TABLECAPTIONS}{TABLECAPTIONS}}\list
	{Table \arabic{enumi}:\hfill}{\settowidth\labelwidth{Table 999:}
	\leftmargin\labelwidth
	\advance\leftmargin\labelsep\usecounter{enumi}}}
 \relax
\def\reflist{\section*{References\markboth
	{REFLIST}{REFLIST}}\list
	{[\arabic{enumi}]\hfill}{\settowidth\labelwidth{[999]}
	\leftmargin\labelwidth
	\advance\leftmargin\labelsep\usecounter{enumi}}}
 \relax
\makeatletter
\newcounter{pubctr}
\def\publist{\@ifnextchar[{\@publist}{\@@publist}}
\def\@publist[#1]{\list
	{[\arabic{pubctr}]\hfill}{\settowidth\labelwidth{[999]}
	\leftmargin\labelwidth
	\advance\leftmargin\labelsep
	\@nmbrlisttrue\def\@listctr{pubctr}
	\setcounter{pubctr}{#1}\addtocounter{pubctr}{-1}}}
\def\@@publist{\list
	{[\arabic{pubctr}]\hfill}{\settowidth\labelwidth{[999]}
	\leftmargin\labelwidth
	\advance\leftmargin\labelsep
	\@nmbrlisttrue\def\@listctr{pubctr}}}
 \relax
\makeatother
\newskip\humongous \humongous=0pt plus 1000pt minus 1000pt

\newif\ifdtup

\relax

\def\thefootnote{\fnsymbol{footnote}}
\def\ref#1{$^{#1)}$}
\def\square{\hbox{{$\sqcup$}\llap{$\sqcap$}}}   
\def\th{\vartheta}
\def\p{\partial}
\def\pb{{\bar \partial}}

\def\k{\kappa}
\def\e{\tilde E}
\def\u{\underline}

\def\a{\alpha}
\def\b{\beta}
\def\g{\gamma}
\def\s{\sigma}

\def\mb{{\bar m}}
\def\qb{{\bar q}}
\def\nb{{\bar n}}
\def\va{{\vec\alpha}}
\def\vb{{\vec\beta}}
\def\t{\theta}
\begin{document}
\begin{titlepage}
\begin{center}
\hfill CERN-TH.6797/93\\
\hfill hep-th/9302033\\
\vskip 1in

{\large \bf Exact Duality Symmetries in CFT and String Theory}

\vskip .8in

\u{Elias Kiritsis}\footnote{email: KIRITSIS@NXTH08.CERN.CH}\\
\vskip .1in

{\em  Theory Division, CERN, CH-1211\\
      Geneva 23, Switzerland}
\end{center}
\vskip .8in

\begin{abstract}
The duality symmetries of WZW and coset models are discussed.
The exact underlying symmetry responsible for semiclassical duality
is identified with the symmetry under affine Weyl transformations.
This identification unifies the treatement of duality symmetries
and shows that in the compact and unitary case they are exact symmetries
of string theory to all orders in $\alpha'$ and in the string coupling
constant.
Non-compact WZW models and cosets are also discussed.
A toy model is analyzed suggesting that duality can survive as an exact
symmetry.
\end{abstract}
\vfill
CERN-TH.6797/93\\
February 1993\\
\end{titlepage}
\newpage
\renewcommand{\thepage}{\arabic{page}}
\setcounter{page}{1}
\section{Introduction, Results and Conclusions}

Strings, being extended objects, sense the target space, into which they are
embended, in a different way than point particles.
The difference comes because, strings,
embended in a compact space, except from their local excitations, that
mimic point particle behaviour (``momentum" modes), have ``winding"
excitations where the string wraps around non-contractible cycles of the
manifold.
The masses of momentum modes  are inversely proportional to the volume of the
manifold, whereas those of the winding modes are proportional to the volume,
since it costs energy in order to stretch the string.
In the simplest possible example, that of a string moving on a circle,
it was observed that the spectrum of the theory with radius $R$ and that with
radius $1/R$ are identical, \cite{dual}.
This duality symmetry is the same as the electric-magnetic duality symmetry
of the underlying 2-d gausian model.
Such duality symetries persist in all flat compact backgrounds, \cite{grv}
and imply the existence of discrete symmetries for the effective theory
of string theory around such backgrounds.
These discrete symmetries are
local, in the sense that they can be considerent as remnants of broken
gauge symmetries, present at special points in the space of such flat
backgrounds, \cite{gau}.

The existence of such symmetries poses important questions about
the background interpretation of such string ground states (CFTs).
Obviously, the string senses the geometry of the target space in a rather
``confusing" way. For example, when the string moves on a circle of radius
R, just looking at the scattering data, we cannot tell if the radius is $R$
or $1/R$.
When $R$ is large or small, then the distinction of the momentum
and winding modes makes sense (although which is which depends on
whether $R$ is large or small).
For $R\sim {\cal O}(1)$ however, such a distinction does not make sense any
more.

To be more specific, we will discuss here the $R\rightarrow 1/R$
duality of a free scalar field in order to set the notation and to derive
the  formula that will be of use for all semiclassical $\s$-model duality
symmetries, \cite{k}.
Consider a scalar field $\phi$ taking values on a circle of radius
$R$. We will use the convention that the $R$ dependence is explicit
and $\phi\in [0,2\pi)$.
Let's consider the partition function in the presence of an external
current $J_{\mu}$
$$Z_{R}(J)=\int_{0}^{2\pi}[RD\phi]exp\left[-{R^{2}\over 4\pi}\int\p_{\mu}
\phi\p^{\mu}\phi+\int\p_{\mu}\phi J^{\mu}\right].\eqno(1.1)$$
In order to perform the duality transformation, we will use an infinite
dimensional version of the gaussian integration formula,
$$e^{-ab^{2}}={1\over 2\sqrt{\pi a}}\int_{-\infty}^{+\infty}dxe^{-{x^{2}\over
4a}+ibx}\eqno(1.2)$$
in order to make the exponent in (1.1) linear in $\phi$.
Thus, we obtain
$$Z_{R}(J)=\int_{0}^{2\pi}[RD\phi]\int\left[{DB_{\mu}\over R^{2}}\right]
exp\left[-{\pi\over R^{2}}\int B_{\mu}B^{\mu}+i\int
B_{\mu}\left(\p^{\mu}\phi-{2\pi\over R^{2}}J^{\mu}\right)+{\pi\over R^{2}}\int
J_{\mu}J^{\mu}\right].\eqno(1.3)$$
The crucial step is to go from the (dummy) vector field $B_{\mu}$ to its dual,
$B_{\mu}=\varepsilon_{\mu\nu}A^{\nu}$.
By integrating out $\phi$ we obtain
$$Z_{R}(J)=\int\left[{2\pi\over
R^{2}}DA_{\mu}\right]\delta(F(A))exp\left[-{\pi\over R^{2}}\int
A_{\mu}A^{\mu}-{2\pi i\over
R^{2}}\int\varepsilon^{\mu\nu}J_{\mu}A_{\nu}+{\pi\over R^{2}}\int
J_{\mu}J^{\mu}\right],\eqno(1.4)$$
where $F(A)=\varepsilon^{\mu\nu}\p_{\mu}A_{\nu}$.
The original theory was invariant under translations of $\phi$ by a constant.
This implies that $\int F(A)=0$.
We will subsequently solve the $\delta$-function constraint by
$A_{\mu}=\p_{\mu}\phi/2\pi$ (the jacobian for this is 1) to finally obtain
$$Z_{R}(J)=\int^{2\pi}_{0}\left[{D\phi\over R}\right]exp\left[-{1
\over 4\pi R^{2}}\int\p_{\mu}\phi\p^{\mu}\phi-{i\over R^{2}}
\int\varepsilon^{\mu\nu}J_{\mu}\p_{\nu}\phi+{\pi\over R^{2}}\int
J_{\mu}J^{\mu}\right].\eqno(1.5)$$
Eq. (1.5) will be enough to derive all $\s$-model duality transformations.
In particular, setting $J=0$, we obtain the usual duality symmetry
$Z_{R}=Z_{1/R}$.\footnote{Similar results can be obtained for correlation
functions. In the path integral framework, the general
conformal operator
(affine U(1) primary) with $(\Delta, {\bar \Delta})=((mR+nR^{-1})^{2}/4,
(mR-nR^{-1})^{2}/4)$ is represented by the insertion of the field
$e^{in\phi(z_{0},{\bar z}_{0})}$ and the instruction to do the
integration over maps $\phi(z,{\bar z})$ which wind $m$
times around the point $z_{0}$.}

The discussion above generalizes to strings propagating on an d-dimensional
torus, where there are d generating duality transformations, each for every
coordinate.

In order to apply (1.5) to a general $\s$-model, the presence of a
Killing symmetry is needed.
In the appropriate coordinates, one can write the action of such a $\s$-model
as
$$S={1\over 4\pi}\int \left[G_{ij}\p_{\mu}x^{i}\p^{\mu}x^{j}+
iB_{ij}\varepsilon^{\mu\nu}\p_{\mu}x^{i}\p_{\nu}x^{j}\right]
,\eqno(1.6)$$
where we assume that $G_{ij},B_{ij}$ do not depend on the coordinate
$x^{0}$.
In terms of $x^{0}$ the action (1.6) has the same form as in (1.1) (we will
assume here that $G_{00}$ is a constant although this is not
necessary\footnote{Non-constant, but $x^{0}$-independent $G_{00}$ can be
handled by the quotient method, \cite{vr,gr}.}).
The identifications are $R^{2}\rightarrow G_{00}$ and
$$J_{\mu}\rightarrow -{1\over 2\pi}\left(G_{0i}\p_{\mu}x^{i}+B_{0i}
\varepsilon_{\mu\nu}\p^{\nu}x^{i}\right).\eqno(1.7)$$
Then application of (1.5) gives a dual action with
$${\tilde G_{00}}={1\over G_{00}}\;\;,\;\;{\tilde G_{0i}}={B_{0i}
\over G_{00}}\;\;,\;\;{\tilde B_{0i}}={G_{0i}\over G_{00}}\eqno(1.8a)$$
$${\tilde G_{ij}}=G_{ij}+{B_{0i}B_{0j}-G_{0i}G_{0j}\over G_{00}}
\;\;,\;\;{\tilde B_{ij}}=B_{ij}+{B_{0i}G_{0j}-B_{0j}G_{0i}\over
G_{00}}.\eqno(1.8b)$$
There is  a change also in the measure, as in (1.5), which can be interpeted as
a shift of the dilaton, (see for example \cite{bu,k,sch}).

In a $\s$-model with $d$ Killing symmetries, the structure of the group
of duality transformations is as follows.
There are $d$ generating duality transformations $D_{i}$, corresponding
to doing the transformation (1.8) in the i-th Killing direction.
These transformations are commutative
$$D_{i}D_{j}=D_{j}D_{i}\eqno(1.9a)$$
and each one generates a $Z_{2}$ group,
$$D_{i}D_{i}=1\;\;,\;\; \forall i=1,2,\cdots,d\eqno(1.9b)$$
The transformation $\prod_{i=1}^{d}D_{i}$ inverts the generalized
metric $G+B\rightarrow (G+B)^{-1}$.

When the target space is a $d$-torus, the $\s$-model is described by (1.6)
with $G,B$ constants.
The partition function can be calculated directly via instanton sums
$$Z=\left({\sqrt{Im\tau}\over |\eta|^{2}}\right)^{d}\sum_{{\vec m},{\vec
n}}e^{-{\pi\over Im\tau}(\tau{\vec m}+{\vec n})_{i}(G+B)_{ij}({\bar \tau}{\vec
m}+{\vec n})_{j}}\;\;.\eqno(1.10)$$
Modular invariance is obvious in (1.10).
Let us introduce the $2d\times 2d$ matrices
$$M=\left(\matrix{G^{-1}&-G^{-1}B\cr BG^{-1}&G-BG^{-1}B\cr}\right)\;\;,\;\;
H=\left(\matrix{0&1\cr 1&0\cr}\right)\;.\eqno(1.11)$$
They satisfy
$$M=M^{T}\;\;,\;\;M^{-1}=HMH\;\;,\;\; M^{T}HM=H\;,\eqno(1.12)$$
which imply that $M\in O(d,d,R)$.
$O(d,d,R)$ transformations act implicitly on $G,B$ via
$M\rightarrow \Omega M \Omega^{T}$.
Upon Poisson resumming (1.10), it can be cast in character form
$$Z={det(G+B)^{-d/2}\over |\eta|^{2d}}\sum_{{\vec m},{\vec n}}q^{\Delta_{
{\vec m},{\vec n}}}{\bar q}^{{\bar \Delta}_{{\vec m},{\vec n}}}\;.\eqno(1.13)$$
Introducing a $2d$ vector ${\vec N}\sim \left(\matrix{{\vec n}\cr {\vec
m}\cr}\right)$, we can write the conformal weights as
$$\Delta_{{\vec m},{\vec
n}}=Q_{i}(G^{-1})_{ij}Q_{j}=N_{i}(M+H)_{ij}N_{j}\;,\eqno(1.14a)$$
$${\bar \Delta}_{{\vec m},{\vec n}}={\bar Q}_{i}(G^{-1})_{ij}{\bar Q}_{j}=
N_{i}(M-H)_{ij}N_{j}\;,\eqno(1.14b)$$
$$Q_{i}=n_{i}+(G-B)_{ij}m_{j}\;\;,\;\;{\bar
Q}_{i}=n_{i}-(G+B)_{ij}m_{j}\;.\eqno(1.14c)$$
The generating duality transformations can be represented as $O(d,d)$
transformations
$$D_{i}\rightarrow \left(\matrix{1-e_{i}&e_{i}\cr
e_{i}&1-e_{i}\cr}\right)\;,\eqno(1.15)$$
where $e_{i}$ is a $d\times d$ matrix with all elements zero except the
diagonal
$ii$ element being $1$.
It is obvious from (1.14) that $D_{i}$ interchanges $m_{i}\leftrightarrow
n_{i}$ and thus the invariance of the partition function is obvious.
In the basis for the currents in which the metric is unity, this amounts
to the transformation ${\bar J}^{i}\rightarrow -{\bar J}^{i}$.

Provided that $J_{\mu}$ is a classical source, eq. (1.5) is exact.
However, subtleties can (and do) arise when $J_{\mu}$ depends on other quantum
fields, as is the case in (1.7).
Of course our semiclassical considerations will still be valid,
but, in general, we expect discrepancies in higher loops.
One of our tasks in this paper is to investigate when semiclassical
duality is exact (in the sense that it can be corrected beyond 1-loop
to yield a genuine symmetry).
There are two points of view relevant here.
One is the $\s$-model point of view, which has the advantage that
the background interpretation is manifest.
The other is the CFT point of view, where, although the background
interpretation is not always obvious, it has the advantage that one can get
exact results easier.

In this paper we will discuss all the duality symmetries of
well-understood CFTs, that is WZW models \cite{wit} and their cosets
\cite{c1,c2}.\footnote{There is a more
general class of CFTs whose structure is much less understood, namely
the affine-Virasoro constructions, \cite{hk}.}
This class is quite large and contains (modulo a mild assumption) all the CFTs
which describe string propagation in a target space with $d$ Killing
symmetries.
It was argued in \cite{gr} that any such $\s$-model can be obtained
by gauging $d$ abelian currents in a WZW model.

The first step is to understand duality in the WZW model.
{}From the $\s$-model point of view, there are many semiclassical duality
transformations, of the type (1.8).
By analyzing their effect on the affine primaries, we will be able to
identify them with Weyl transformations acting on the current
algebra representations.
The question of exact duality invariance then translates into
invariance under the affine Weyl group.
In the case of compact current algebra and unitary (integrable)
representations the affine Weyl group is a genuine symmetry.
This is not the case
in general (where it relates inequivalent representations).
We will also see explicitly that the action of the exact duality
transformation on the fields is, in general, more complicated than the
semiclassical duality transformation.

Once we understand how duality works in the WZW model, we can proceed
to the coset models. When we gauge a semisimple subgroup, then the
duality symmetry of the coset theory is inherited from that of the original
WZW model, and the different dual actions are obtained by gauging the
different dual actions of the WZW model. The non-trivial duality
transformations are those that leave the subgroup structure invariant.
The generic coset $G/H$ model with H semisimple, has extra Killing symmetries,
which can be used to generate duality transformations. However
these transformations are included in the ones mentioned above.
When $H$ is maximal, then the $\s$-model describing the $G/H$ coset
has no Killing symmetries. However, according to our previous discussion
it still posseses duality symmetries.

More interesting things happen when H is abelian.
In this case, we have the option to gauge an axial or a vector abelian
current.
Semiclassically, it can be shown that, these two theories are dual
to each other, \cite{k,dvv}.
We will see that the original affine Weyl symmetry of the WZW model
guarantees that this extra axial-vector duality is an exact symmetry
(although in the $\s$-model language it needs corrections beyond one-loop ).
This type of duality is a generalization of the order-disorder
(Kramers-Wannier) duality of the critical Ising model.
Using axial-vector duality, one can generate new conformal
$\s$-models using $O(d,d,R)$ transformations, \cite{ven,sen,js}.
The $O(d,d,R)$ transformations need to be corrected
beyong one-loop, however, in the compact case, this can always be done.
One implication of this result is that there are marginal
$J{\bar J}$ perturbations in $\s$ models with Killing symmetries.
If the currents are abelian and chiral this is already known.
However marginality persists for some combinations of non-chiral
abelian currents.

The presence of duality symmetries in compact targets complicates
the background interpretation of the $\s$ model.
When $\s$-model couplings are strong, it is difficult to
have a geometric notion of a target manifold (even the notion
of dimensionality can break down, and many such instances are known, for
example
$SU(2)_{k=1}\sim U(1)_{R=1}$ etc.).
The only case where one has an (almost) unabiguous notion of a manifold
is when all couplings are weak, ($\a '\rightarrow 0$).
In curved backgrounds, the dual versions obtained for example by (1.8)
are not trustworthy guides of geometry since the dual background
describes strong coupling regions.

When one considers string propagation in non-compact backgrounds, the
situation could be quite different.
For Euclidean non-compact cosets,
although the underlying affine Weyl group
relates, in general, inequivalent representations we will give indications
that duality symmetry will be preserved. A potential reason for survival
of duality here might be that the
spectrum can be classified into complete orbits of
the affine Weyl group.\footnote{This has been
effectively done for the SL(2,R)/U(1) coset in \cite{hw}.}
There are two potential problems with this procedure. The first is
that the required orbits contain representations that are not positive.
However, this might not be lethal for the associated string model, but
positivity of the string Hilbert space needs to be addressed.
\footnote{In the case of $SL(2,R)$ this orbit method works for the discrete
series but it is not at all obvious how it could be implemented in the
continuous series.}
The second is that the background interpretation of such theories is obscure.
We should stress here though that we have no evidence, that this mechanism
is the one responsible for the survival of duality.

In order to investigate whether the semiclassical duality is exact in the
non-compact case we will analyse the simplest possible model, where
axial-vector
duality relates the 2-d Euclidean plane (free field theory), to a certain
singular manifold.
Although we cannot compute the latter partition function exactly, we will
compute it in an improved ``minisuperspace" approximation where it will turn
out
to be equal to that of the plane.
Although, this computation does not settle the issue of exactness of
non-compact duality symmetries it does give some useful indications.

For non-compact cosets with Minkowskian signature, the meaning of the
duality transformation is different.
Instead of relating two different manifolds, it interchanges various regions of
spacetime, \cite{giv,dvv}.
The same remarks apply here as in the Euclidean case.
Duality here, although it might not be a symmetry, provides a map
that can give meaning to regions of spacetime that one otherwise would
traditionally neglect.

The structure of the rest of this paper is as follows. In section 2 we will
analyze semiclassically and exactly the duality symmetries of compact WZW
models.
The same will be done for compact cosets in section 3.
The extension of duality symmetries to $O(d,d)$ symmetries will be discussed
in section 4. Section 5 contains some remarks on marginal current-current
perturbations implied by $O(d,d)$ covariance.
Finally, in section 6 we will discuss non-compact cosets.

\section{Duality in the WZW model.}

In this section we will analyze in detail the duality symmetries of WZW model,
both from the $\s$-model and the CFT (affine current algebra) point of view.

We will consider for simplicity a compact group $G$ which is simple
and simply laced. It will turn out that understanding the simplest
such group, SU(2), will suffice. In the case of non-simply laced simple groups
there are some minor changes due to the short roots that will be dealt with
latter on.
The case of non-simple groups has further complications that we will
not consider here.

The action of the WZW model is
$$I(g)={k\over 4\pi}I_{NS}(g)+{ik\over 6\pi}\Gamma_{WZ}(g)\eqno(2.1)$$
$$I_{NS}(g)=\int d^{2}x Tr[U_{\mu}U^{\mu}]\;\;,\;\;\Gamma_{WZ}(g)=\int
\limits_{B\atop \p B=S^{2}}
d^{3}y\varepsilon^{\mu\nu\rho}Tr[U_{\mu}U_{\nu}U_{\rho}]\eqno(2.2)$$
where
$$U_{\mu}=g^{-1}\p_{\mu}g\;\;,\;\;V_{\mu}=\p_{\mu}g g^{-1}\eqno(2.3)$$
$g$ is a matrix in the fundamental representation of $G$ and $Tr$ is
a properly normalized trace such that
$${1\over 12\pi^{2}}\int_{S^{3}}Tr[U\wedge U\wedge U]\in Z\;.\eqno(2.4)$$
The action $I(g)$ is invariant under the group $G_{R}\otimes G_{L}$,
generated by left and right group transformations, $g\rightarrow h_{1}gh_{2}$,
with associated conserved currents
$$J^{\mu}_{R}={k\over 2\pi}P_{-}^{\mu\nu}U_{\nu}\;\;,\;\;J^{\mu}_{L}=
{k\over 2\pi}P_{+}^{\mu\nu}V_{\nu}\eqno(2.5)$$
with $P_{\pm}^{\mu\nu}\equiv\delta^{\mu\nu}\pm i\varepsilon^{\mu\nu}$.
These currents are conserved and chirally conserved and they generate
two copies of the affine $\hat G$ current algebra.
An important property of the WZW action is that it satisfies the
Polyakov-Wiegman formula
$$I(gh)=I(g)+I(h)-{k\over 2\pi}\int
d^{2}xP^{\mu\nu}_{+}Tr[U_{\mu}(g)V_{\nu}(h)]\eqno(2.6)$$

To generate duality transformations in the WZW model, we pick a
generator of the Lie algebra of $G$, $T^{0}$, normalized as
$Tr[(T^{0})^{2}]=1$.
We can then parametrize $g=e^{i\phi T^{0}}h$.
Using (1.6), the action $I(g)$ takes the form
$$I(g)=I(h)+{k\over 4\pi}\int \p_{\mu}\phi\p^{\mu}\phi -{ik\over 2\pi}
\int P^{\mu\nu}_{+}\p_{\mu}\phi V^{0}_{\nu}(h)\eqno(2.7)$$
where $V^{0}_{\mu}(h)=Tr[T^{0}V_{\mu}(h)]$.
We can now apply the duality map (1.5) $\rightarrow$ (2.7) to obtain\footnote{
The measure also changes by a finite computable piece, see \cite{k}}
$$I^{\rm dual}(g)=I(h)+{1\over 4\pi k}\int \p_{\mu}\phi\p^{\mu}\phi-
{i\over 2\pi}\int P_{+}^{\mu\nu}\p_{\mu}V^{0}_{\nu}(h)\;.\eqno(2.8)$$
The angle $\phi$ was originally normalized to take values in $[0,2\pi]$.
It is obvious from (2.8) that the effect of the duality transformation
is to change the range of values to $[0,2\pi/k]$.
To see how many independent duality transformations exist, we have to
explicitly parametrize the Cartan torus dependence of the WZW model.
Pick a basis in the Cartan algebra, $T^{i}$, $i=1,2,\cdots,r$,
$[T^{i},T^{j}]=0$, $Tr[T^{i}T^{j}]=\delta^{ij}$ and parametrize,
$$g=e^{i\sum_{i=1}^{r}\a^{i}T^{i}}\,h\,e^{i\sum_{i=1}^{r}\g^{i}T^{i}}
\;\;.\eqno(2.9)$$
Then using (2.6) the WZW action becomes
$$I(g)=I(h)+{k\over 4\pi}\int(\p_{\mu}\a^{i}\p^{\mu}\a^{i}+\p_{\mu}\g^{i}
\p^{\mu}\g^{i})-{ik\over 2\pi}\int(P_{+}^{\mu\nu}\p_{\mu}\a^{i}V^{i}_{\nu}
(h)+P_{-}^{\mu\nu}\p_{\mu}\g^{i}U_{\nu}^{i}(h))+$$
$$+{k\over 2\pi}\int
P_{+}^{\mu\nu}\p_{\mu}\a^{i}\p_{\nu}\g^{j}M^{ij}(h)\eqno(2.10)$$
where
$$U_{\mu}^{i}(h)=Tr[T^{i}U_{\mu}(h)]\;,\;V_{\mu}^{i}(h)=Tr[T^{i}V_{\mu}(h)]
\;,\; M^{ij}(h)=Tr[T^{i}hT^{j}h^{-1}]\;.\eqno(2.11)$$
It is obvious from (2.10) that we can apply the duality transformation
using  any of the $\a^{i}$, $\g^{i}$.
Thus, there are $2^{2r}-1$ non-trivial duality transformations.
A duality transformation on $\a^{i}$ effectively makes the substitution
$\a^{i}\rightarrow \a^{i}/k$ in the action whereas a duality transformation
on $\g^{i}$ makes the substitution $\g^{i}\rightarrow -\g^{i}/k$.

In order to identify the underlying property of the WZW model, responsible
for the invariance under these duality transformations, we have delve a bit
into such elements of the the representation theory of the affine Lie algebras
as the affine Weyl group and external automorphisms.
Here, I will just state some properties that we need. More information
can be obtained in \cite{gw} and references therein.
\setcounter{footnote}{0}

The affine Weyl group ${\hat W}$ is a semidirect product of the Lie algebra
Weyl group $W$ times a translation group, ${\hat W}=W\triangleright T$.
Appart from the action of finite Weyl group elements, there are Weyl
transformations associated to roots which have a component in the direction of
the imaginary simple root.
The action of such an element ${\hat W}_{\vec\a}$
on a finite Lie algebra weight $\vec\lambda$ and on the grade $n$ is
$${\hat W}_{\vec\a}({\vec\lambda})=W_{\vec\a}({\vec\lambda})-k{\vec\beta}
\eqno(2.12a)$$
$${\hat W}_{\vec\a}(n)=n-{\vec\lambda}\cdot{\vec\beta}-{k\over
2}{\vec\beta}\cdot{\vec\beta}\eqno(2.12b)$$
where ${\vec\beta}=2{\vec\a}/{\vec\a}\cdot{\vec\a}$ is the coroot associated
to the finite Lie algebra root $\vec\a$, the grade $n$ is
basically the mode number\footnote{In a highest weight representation where
the affine primaries have $L_{0}$ eigenvalue $\Delta$, the grade $n$ of a
state is the eigenvalue of $L_{0}-\Delta$ on that state.} and $W_{\vec\a}
(\vec\lambda)={\vec\lambda}-{\vec\a}({\vec\lambda}\cdot{\vec\beta})$ is a
finite Weyl transformation.
It is important to note that affine Weyl transformations, in general, map
states inside a representation at different levels.

There are also external automorphisms of the affine algebra which are
essentially associated to symmetries of the affine Dynkin diagram.
For the $SU(n)$ case, the affine Dynkin diagram consists of $n$ nodes
connected around a circle. The external automorphisms are generated
by a basic rotation, and a reflection which corresponds to the finite Lie
algebra external automorphism (that maps a representation to its complex
conjugate). When we write a highest weight ${\vec\Lambda}=\sum_{i=1}^{n-1}
m_{i}{\vec\Lambda}_{i}$ in terms of the fundamental weights
${\vec\Lambda}_{i}$, ($m_{i}$ are non-negative integers), the action of the
generating rotation of the affine Dynkin diagram is as follows
$$\sigma({\vec\Lambda})=(k-\sum_{i=1}^{n-1}m_{i}){\vec\Lambda}_{1}+
m_{1}{\vec\Lambda}_{2}+\cdots+m_{n-2}{\vec\Lambda}_{n-1}\;.\eqno(2.13)$$
$\sigma$ generates a $Z_{n}$ group\footnote{In general this group is isomorphic
to the center of the finite Lie group} where $\sigma^{n}=1$ on the
heighest weights, but acts as an affine Weyl transformation in the
representation.
Specializing to SU(2), let $m\in Z/2$ be the weight, and $j\in Z/2$
the highest weight (spin of a representation).
Then the finite Weyl group acts as $m\rightarrow -m$, and combined with
the affine translation $m\rightarrow m+k$ they generate the affine Weyl group.
The only nontrivial outer automorphism $\sigma$ acts as $j\rightarrow k-j$
and $\sigma^{2}$ is a Weyl translation.

The non-trivial statement now is: For compact groups, integer level
and integrable heighest weight representations, both the affine Weyl group
and the external automorphisms are symmetries.
In particular, in a WZW model the Hilbert space is constructed by tying
together (in a modular invariant way) two copies of representations of the
affine algebra. Thus, we have invariance under independent affine Weyl
transformations acting on left or right representations.
Moreover, since the modular transformation properties of the affine characters
reflect the external automorphism symmetries, the theory is invariant
under external automorphisms that act at the same time on left and right
representations.
These invariance properties can be verified for correlation functions on the
sphere and the torus. This then implies that they hold on an arbitrary Riemann
surface
since the sphere and torus data are sufficient in order to construct
the correlators at higher genus.

As an example, we will present the SU(2) case and focus on the spectrum.
We introduce the (affine) SU(2)$_{k}$ characters
$$\chi_{l}(q=e^{2\pi i\tau}, w)=Tr_{l}\left[q^{L_{0}}e^{2\pi iwJ^{3}_{0}}
\right]=\sum_{m=-k+1}^{k} c^{l}_{m}(q)\vartheta_{m,k}(q,w)\eqno(2.14)$$
where $l$ is twice the spin (a non-negative integer) and  $m$ is twice the $
J^{3}_{0}$ eigenvalue.
The trace is in the affine hw representation of spin $l$,
$$\vartheta_{m,k}(q,w)=\sum_{n\in Z} q^{k(n+{m\over 2k})^{2}}e^{2\pi
iw(kn+{m\over
2})}\eqno(2.15)$$
and $c^{l}_{m}$ are the standard string functions \cite{kp} which satisfy
$c^{l}_{m}=0$ when $l-m=1(mod$ $2)$ (which means that the spin is increased
or decreased in units of 1) .
For integrable representations ($k$ is a positive integer and $0\leq l\leq k$),
invariance under the affine Weyl group is equivalent to
$$c^{l}_{m}=c^{l}_{-m}\;\;,\;\; c^{l}_{m}=c^{l}_{m+2k}\eqno(2.16)$$
The first relation is due to the Weyl group of $SU(2)$ while the second is the
generating translation in the affine Weyl group.
There is another important relation
$$c^{l}_{m}=c^{k-l}_{k-m}\eqno(2.17)$$
which is a consequence of the external affine automorphism, \cite{kp}.

The duality tranformation on $\a^{i}$ amounts to replacing ${\bar J}^{i}
\rightarrow -{\bar J}^{i}$, where ${\bar J}^{i}$ is the right Cartan current
in the $T^{i}$ basis of the Cartan subalgebra.
Similarly the duality transformation on $\g^{i}$ amounts to the replacement
$J^{i}\rightarrow -J^{i}$ at the level of the Cartan subalgebra.
This is not the whole story however. With a bit more effort one can see that
they
act as Weyl transformations on the left or right SU(2) currents.
This identification can be seen clearly by coupling the WZW action to external
gauge fields and monitoring the effect of the duality transformation on the
currents. It can also be recovered  from the twisted partition function
via the action of the duality transformation on the gauge field moduli
(for the Cartan).
Let us now check that the duality transformations
$$D_{i}\;\;:\;\; J^{i}\rightarrow -J^{i}\eqno(2.18a)$$
$${\bar D}_{i}\;\;:\;\;{\bar J}^{i}\rightarrow -{\bar J}^{i}\eqno(2.18b)$$
are exact symmetries of the model.
We will consider again for simplicity SU(2) and then generalize to an
arbitrary group.

The partition function is
$$Z(q,{\bar q})=\sum_{l,{\bar l}=0}^{k}N^{l,{\bar l}}\chi_{l}(q,w=0){\bar
\chi}_{\bar l}({\bar q},{\bar w}=0)\eqno(2.19)$$
where $N^{l,{\bar l}}$ is one of the CIZ modular invariants. The diagonal
one $N^{l,{\bar l}}=\delta_{l,{\bar l}}$ corresponds to the usual WZW
model.
Putting everything together we obtain
$$Z(q,{\bar q})=\sum_{l,{\bar l}=0}^{k}\sum_{m=-k+1}^{k}
\sum_{\mb =-k+1}^{k}N^{l,{\bar l}}c_{m}^{l}(q){\bar c}^{\bar l}_{\mb}(\qb)
\cdot$$
$$\cdot \sum_{n,\nb \in Z}exp\left[2\pi i\left(\tau k(n+{m\over
2k})^{2}-{\bar \tau}k(\nb +{\mb \over 2k})^{2}\right)\right]\eqno(2.20)$$
The two generating duality transformations here correspond to $n\rightarrow
-n$, $m\rightarrow -m$, and ${\bar n}\rightarrow -{\bar n}$, ${\bar
m}\rightarrow -{\bar m}$.
They are symmetries of (2.20) if we use the invariance of the string functions
under the affine Weyl group, (2.16).

This invariance is similar, but qualitatively different than that present
in flat backgrounds. There, one has a family of theories parametrized
by $G,B$ and duality is the statement that two theories are equivalent
for different values of the parameters. Here, there is no parameter present
and, in this sense, this is what we could call self-duality.
This becomes more transparent if we consider the one parameter family of
theories, parametrized by the radius of the cartan torus of SU(2).
The partition function is known, \cite{sky}
$$Z(R)=\sum_{l,{\bar l}=0}^{k}\sum_{m=-k+1}^{k}\sum_{r=0}^{k-1}N^{l,{\bar
l}}c^{l}_{m}(q){\bar c}^{\bar l}_{m-2r}(\qb)\sum_{M,N\in
Z}q^{\Delta_{M,N}}{\bar
q}^{{\bar \Delta}_{M,N}}\eqno(2.21)$$
with
$$\Delta_{M,N}={1\over 4k}\left({kM+m-r\over R}+R(kN+r)\right)^{2}\;\;
,\;\,{\bar \Delta}_{M,N}={1\over 4k}\left({kM+m-r\over
R}-R(kN+r)\right)^{2}\eqno(2.22)$$
In (2.21) there is a duality symmetry $R\rightarrow 1/R$, which becomes
self-duality at the point $R=1$, that corresponds to the WZW
model.

Now we are in a position to discuss the general WZW model for a simple
group $G$. Let $M$ be the root lattice, $M_{L}$ the long root lattice
and $M^{*}$ the weight lattice.
The character of a hw representation of ${\hat G}$ with hw $\vec \Lambda$
is defined as
$$\chi_{\vec \Lambda}(q,{\vec w})=Tr[q^{L_{0}}e^{2\pi i{\vec w}\cdot{\vec
J}_{0}}]\eqno(2.23)$$
where ${\vec J}_{0}$ generates the cartan subalgebra of $G$.
The character admits the string function decomposition, \cite{kp}
$$\chi_{\vec \Lambda}=\sum_{{\vec \lambda}\in M^{*}/kM_{L}} c^{\vec
\Lambda}_{\vec \lambda}(q)\Theta_{\vec \lambda}({\vec w},q)\eqno(2.24)$$
with $\Theta_{\vec\lambda}$ being the classical $\vartheta$-function
of level $k$ of the Lie algebra of $G$
$$\Theta_{\vec\lambda}({\vec w},q)=\sum_{{\vec\gamma}\in M_{L}}
q^{{k\over 2}\left({\vec\gamma}+{{\vec\lambda}\over k}\right)^{2}}
e^{2\pi i{\vec w}\cdot(k{\vec\gamma}+{\vec\lambda})}\;.\eqno(2.25)$$
The string functions are invariant under the Weyl group and Weyl translations
$$c^{\vec\Lambda}_{w({\vec\lambda})}=c^{\vec\Lambda}_{\vec\lambda}\;\;,\;\;
c^{\vec\Lambda}_{{\vec\lambda}+k{\vec\beta}}=c^{\vec\Lambda}_{\vec\lambda}
\eqno(2.26)$$
where $w$ is a Weyl transformation and ${\vec\beta}\in M_{L}$.

The (left) generating duality transformations $D_{i}$ correspond to Weyl
reflections generated by the simple roots ${\vec\a}_{i}$ which implement
the transformations (2.18a).
The invariance of the spectrum (and partition function) is encoded in the
fact, obvious from (2.25,26), that $\chi_{\vec\Lambda}$ is invariant under
$w_{i}\rightarrow -w_{i}$.
Although $w_{{\vec\a}_{i}}$ do not commute, they do so when applied to the
character, thus at the level of the partition function they generate a group
isomorphic to (1.9).
However, at the level of correlation functions the (left) duality group is
larger and in fact isomorphic to the finite Weyl group of $G$, $W_{G}$.
Thus the full duality group of the WZW model is $W_{G}\times W_{G}$
the first acting on the left current modules while the second acting
on the right current modules.\footnote{In \cite{dq} a non-abelian form of
duality transformations was introduced. It is not clear if these are related
to the extended duality group introduced above.}
The structure of the (self)-duality group is different than the one present
in flat backgrounds.

\section{Compact Cosets}

A host of CFTs can be obtained from the coset costruction \cite{c1}.
In Langrangian form it amounts to gauging a subgroup $H$ of $G$ in a
conformally invariant way, \cite{c2}.
\setcounter{footnote}{0}
We will assume $G$ to be simple, and $H$ regularly embedded\footnote{The
analysis can be extended to non-simple $G$ and/or irregularly embedded $H$, but
it is certainely more involved.}.

Let us first consider $H$ to be semi-simple. Then, there is one possible
gauging
, the vectorial one, \cite{k}. The gauged WZW action is
$$S_{V}(g,A)=I(g)+{k\over 2\pi}\int
d^{2}xTr[(J^{\mu}_{R}-J^{\mu}_{L})A_{\mu}+P_{+}^{\mu\nu}A_{\mu}gA_{\nu}
g^{-1}-A_{\mu}A^{\mu}]\eqno(3.1)$$
where $A_{\mu}$ belongs to the Lie algebra of $H$.
The action (3.1) is invariant under
$$g\rightarrow hgh^{-1}\;\;\;,\;\;\;A_{\mu}\rightarrow
h^{-1}A_{\mu}h+h^{-1}\p_{\mu}h\;.\eqno(3.2)$$
Since the gauge field is quadratic in the action (3.1) one can integrate it
out, and fix a physical gauge in order to obtain a $\s$-model desription
of the coset theory.

There are two possible ways to generate duality transformations for the
non-abelian coset theory.
The first is to gauge different dual versions of the original WZW model.
The group of the left duality transformations obtained this way is equivalent
to the original Weyl group $W_{G}$ with the restriction that its subgroup
$W_{H}$ acts trivially. The action of $W_{H}$ can be absorbed in a
redefinition of the gauge fields, and in the $\s$-model form
(where the gauge fields have be integrated out) is trivial.

The other possibility is that the action (3.1) has Killing symmetries
which can be exploited in order to generate duality transformations.
The constant vector gauge transformations are symmetries of (3.1) but will
not survive the passage (gauge fixing) to the $\s$-model.
We can directly hunt for such Killing symmetries.
The result is that whenever
there exists a subgroup $H'$ of $G$, such that $[H,H']=0$, then, there are
extra conserved currents which can be calculated from (3.1),
$$J^{\mu}_{R}=P_{+}^{\mu\nu}g^{-1}(\p_{\nu}g+\{g,A_{\nu}\})|_{H'}\eqno(3.3a)$$
$$J^{\mu}_{L}=P_{-}^{\mu\nu}(\p_{\nu}g-\{g,A_{\nu}\})g^{-1}|_{H'}\eqno(3.3b)$$
where $\{,\}$ stands for anti-commutator, and $|_{H'}$ implies a projection
onto the Lie algebra of $H'$.
The currents (3.3) transform covariantly under $H$ gauge transformations
and they are conserved: $\p_{\mu}J^{\mu}_{R,L}=0$.
A less obvious, but verifiable statement is that these currents are also
chirally conserved: $\varepsilon_{\mu\nu}\p^{\mu}J^{\nu}_{L,R}=0$.
Thus, they generate a $H'$ current algebra, and this implies that, locally
the $G/H$ model can be factorized into a $G/(H\times H')$ model times a
$H'$ WZW model. This guarantees the presence of the Killing symmetries
associated to the Cartan of $H'$ and the duality transformations they
imply have been discussed in the previous section.

A more interesting case is when $H$ is abelian.
We will assume without much loss of generality that $H=U(1)$. The situation
with more $U(1)$'s will become obvious.
In this case, there are two possible ways to gauge. The vector as in the
non-abelian case with action given in (3.1) and the axial with action
$$S_{A}(g,A)=I(g)+{k\over 2\pi}\int
d^{2}xTr[(J^{\mu}_{R}+J^{\mu}_{L})A_{\mu}-P_{+}^{\mu\nu}A_{\mu}gA_{\nu}
g^{-1}-A_{\mu}A^{\mu}]\;.\eqno(3.4)$$
The axial and vector actions are related by a duality transformation,
\cite{dvv,k}.
In order to show this, we have to parametrize the group element $g$ as in (2.7)
where $T^{0}$ is the generator of the U(1) subgroup.
In order to write the gauged action (3.1), we need the expressions for the
left and right U(1) currents
$$J^{\mu}_{R}={k\over 2\pi}P_{-}^{\mu\nu}(iX(h)\p_{\nu}\phi+U^{0}_{\nu}
(h))\;\;,\;\;J^{\mu}_{L}={k\over 2\pi}P_{+}^{\mu\nu}(i\p_{\nu}\phi+
V^{0}_{\nu}(h))\eqno(3.5)$$
where $X(h)=Tr[T^{0}hT^{0}h^{-1}]$, as well as (2.7) for the WZW action.
Then,
$$S_{V}(g,A)=I(h)+{k\over 4\pi}\int \p_{\mu}\phi\p^{\mu}\phi -{ik\over 2\pi}
\int P^{\mu\nu}_{+}\p_{\mu}\phi V^{0}_{\nu}(h)+$$
$$+{ik\over 2\pi}\int
A_{\mu}\left(P_{-}^{\mu\nu}(iX(h)\p_{\nu}\phi+U^{0}_{\nu}(h))
-P_{+}^{\mu\nu}(i\p_{\nu}\phi+V^{0}_{\nu}(h))\right)-{k\over 2\pi}\int
(1-X(h))A_{\mu}A^{\mu}.\eqno(3.6)$$
Applying the duality transformation (1.5), we obtain
$$S_{V}\rightarrow S_{V}^{\em dual}=I^{\em dual}(g)+{ik\over 2\pi}\int
A_{\mu}({\tilde J}^{\mu}_{R}+{\tilde J}^{\mu}_{L})-{k\over 2\pi}\int
(1+M)A_{\mu}A^{\mu}\eqno(3.7)$$
where ${\tilde J}^{\mu}_{R,L}$ are the respective currents of the dual theory
$${\tilde J}^{\mu}_{R}={k\over 2\pi}P_{-}^{\mu\nu}\left({i\over k}X(h)
\p_{\nu}\phi+U^{0}_{\nu}(h)\right)\;\;,\;\;{\tilde J}^{\mu}_{L}={k\over
2\pi}P_{+}^{\mu\nu}\left({i\over k}\p_{\nu}\phi+V^{0}_{\nu}(h)\right)
\eqno(3.8)$$
Inspection of (3.7) shows that it is the axially gauged dual WZW model action.
Of course, this is not unexpected, since, at the naive level, the vector coset
is the WZW model with the constraint $J_{L}-J_{R}=0$. As we have seen, a
duality
transformation of the WZW model changes the sign of one of the currents, and
this
gives the axial constraint $J_{L}+J_{R}=0$.

\setcounter{footnote}{0}
In order to investigate to what extend this semiclassical axial-vector
duality is exact, we will analyze the partition function.
In particular we will need a method to compute exactly the partition
function for both the axial and the vector gauge theory.
The easiest way is the operator method.\footnote{More precisely it is a
hybrid of operator
methods and the path integral approach of Gawedski, \cite{c2}.}

We will start by considering the $SU(2)_{k}/U(1)$ coset which captures
the relevant effects. Once we understand it, the generalization will be simple.

Since we are concerned with the partition function, we will be working on
the torus, in the standard flat metric.
It is well known, \cite{c2} that, in the gauge
 $\p_{\mu}A^{\mu}=0$,
the gauged WZW action factorizes (up to gauge field moduli) to that of the
original WZW plus the quadratic action for the gauge field (and the FP
determinant, det$'${\square}).
The effect of the gauge field moduli is to introduce twisted boundary
conditions for the field $g$ of the WZW model around the two non-contactible
cycles of the torus.
The strategy will be to compute the WZW partition function in the presence
of the gauge-field moduli (twists), then integrate over them as specified
by the gauge field measure, and then add the contribution of the (decoupled)
local part of the gauge field.
Consider first the twist in the ``space" direction,
(vector gauging is considered here).
Its effect is to impose a boundary condition on $g$ which is a global
$U(1)_{V}$ transformation ,

$$g(\sigma +1)=e^{i\pi\alpha\s_{3}}g(\s)e^{-i\pi\alpha\s_{3}}\;.\eqno(3.9)$$
The left and right currents are defined in the standard fashion
$$J={k\over 2\pi}g^{-1}\p g\,\,\,,\,\,\,{\bar J}={k\over 2\pi}{\bar \p}g
g^{-1}\;.\eqno(3.10)$$
If we introduce cylinder coordinates
$$z=e^{2\pi(t+i\s)}\;\;,\;\;{\bar z}=e^{2\pi(t-i\s)}\eqno(3.11)$$
(3.9) amounts to
$$J^{\pm}(ze^{2\pi i})=e^{\pm 2\pi i\a}J^{\pm}(z)\;\Rightarrow \;
J^{\pm}(z)=\sum
_{m\in Z\mp \a} {J^{\pm}_{m}\over z^{m+1}}\eqno(3.12a)$$
$${\bar J^{\pm}}({\bar z}e^{-2\pi i})=e^{\pm 2\pi i\a}{\bar J}^{\pm}({\bar z})
\;\Rightarrow \; {\bar J}^{\pm}({\bar z})=\sum _{m\in Z\mp \a}
{{\bar J}^{\pm}_{m} \over {\bar z}^{m+1}}\eqno(3.12b)$$
while it leaves $J^{3},{\bar J}^{3}$ almost invariant. In fact, due to the
central term in the current algebra, both $J^{3},{\bar J}^{3}$ are shifted by
the same constant, to be determined below. The vector current $J^{3}-{\bar
J}^{3}$ is invariant, as it should be.

The twisted currents satisfy an algebra that is isomorphic to the untwisted
SU(2) current algebra.
In particular, the Cartan currents are shifted,
$$J_{m}^{3}(\a)=J^{3}_{m}+{k\a\over 2}\delta_{m,0}\eqno(3.13)$$
and similarly for ${\bar J}^{3}$.
Then,
$$[J^{+}_{m-\a},J^{-}_{n+a}]={k\over
2}m\delta_{m+n,0}+J^{3}_{m+n}(\a)\eqno(3.14a)$$
$$[J^{3}_{m}(\a),J^{\pm}_{n\mp\a}]=\pm J^{\pm}_{m+n\mp\a}\eqno(3.14b)$$
$$[J^{3}_{m}(\a),J^{3}_{n}(\a)]={k\over 2}m\delta_{m+n,0}\eqno(3.14c)$$
and similarly for the left sector.

The Virasoro operator also get shifted.
This is standard, we can see it by either doing the Sugawara construction
using the twisted currents or checking that the following expression has the
proper commutation relations
$$L_{m}(\a)=L_{m}+\a J^{3}_{m}+{\a^{2}k\over 4}\delta_{m,0}\;.\eqno(3.15)$$

Now we need to twist in the ``time" direction. This is achieved in the standard
way by inserting a factor
$$e^{2\pi i\b(J^{3}_{0}(\a)-{\bar J}^{3}_{0}(\a))}\eqno(3.16)$$
which will eventually project onto invariant states (this is similar to the
orbifold case).
Thus, collecting everything together, we obtain the (vectorially) twisted
WZW partition function
$$Z_{\a}^{\b}(q,{\bar q};V)=Tr_{H}\left[ q^{L_{0}(\a)}{\bar q}^{{\bar L}_{0}
(\a)}e^{2\pi i\b(J^{3}_{0}(\a)-{\bar J}^{3}_{0}(\a))}\right]\eqno(3.17)$$
where $H$ is the Hilbert space of the WZW theory.
Using (2.14,15,19) and (3.13,15) we can explicitly evaluate (3.17),
$$Z_{\a}^{\b}(q,{\bar q})=\sum_{l,{\bar l}=0}^{k}\sum_{m=-k+1}^{k}
\sum_{\mb =-k+1}^{k}N^{l,{\bar l}}c_{m}^{l}(q){\bar c}^{\bar l}_{\mb}(\qb)
\cdot$$
$$\cdot \sum_{n,\nb \in Z}exp\left[2\pi i\left(\tau k(n+{m\over 2k}+
{\a\over 2})^{2}-{\bar \tau}k(\nb +{\mb \over 2k}+{\a\over 2})^{2}+
\b(k(n-\nb)+{m-\mb\over 2})\right)\right].\eqno(3.18)$$
 The twisted partition function satisfies
$$Z_{\alpha}^{\beta}=Z^{\beta}_{\alpha+1}=Z^{\beta+1}_{\alpha}=
Z^{-\beta}_{-\alpha}\eqno(3.19)$$
which can be shown, using the invariance under the affine Weyl group, (2.16)
$and$ the invariance under the proper outer automorphism, (2.17).
Eq. (3.19) specifies the fundamental domain for the gauge field moduli, and
agrees with the periodicity implied by the $U(1)$ transformations (3.9,16).
Under modular transformations it transforms as
$$Z^{\beta}_{\alpha}(\tau+1,{\bar \tau}+1)=Z^{\alpha+\beta}_{\alpha}(\tau,{\bar
\tau})\eqno(3.20a)$$
$$Z^{\beta}_{\alpha}(-{1\over \tau},-{1\over {\bar \tau}})
=Z^{-\alpha}_{\beta}(\tau,{\bar \tau})\eqno(3.20b)$$
Eqs. (3.15) imply that, under a modular transformation
$$\tau\rightarrow {a\tau+b\over c\tau+d}\;\;,\;\; \left(\matrix{a&b\cr
c&d\cr}\right)\,\in\, SL(2,Z)\eqno(3.21)$$
the gauge field moduli transform linearly
$$\left(\matrix{\a\cr \b\cr}\right)\rightarrow \left(\matrix{a&b\cr c&d\cr}
\right)\;\left(\matrix{\a\cr \b\cr}\right)\eqno(3.22)$$
The meaning of (3.20-22) becomes more transparent if we introduce complex
coordinates in the gauge field moduli space, $u=\alpha\tau+\beta$.
Then, using (3.19) we can see that the twisted partition function $Z(u,{\bar
u},
\tau,{\bar \tau})$ is invariant under the mapping class group of a torus
with coordinate $u$ and modulus $\tau$,
$$u\rightarrow u+1\,\,\,,\,\,\,u\rightarrow u+\tau\eqno(3.23a)$$
$$\tau\rightarrow\tau+1\,\,\,,\,\,\, u\rightarrow u\eqno(3.23b)$$
$$\tau\rightarrow -{1\over \tau}\,\,\,,\,\,\, u\rightarrow {u\over
\tau}\;.\eqno(3.23c)$$

It remains to calculate the integral over the fundamental region of the
moduli
$$\int_{0}^{1}d\a\int_{0}^{1}d\b Z_{\a}^{\b}(q,\qb;V)\;.$$
We can do the integral over $\b$ first. The only terms in the sum (3.18)
that contribute are those that satisfy $k(n-\nb)+{m-\mb\over 2}=0$
Taking into account the ranges of $m,\mb$ and the fact that they are both
even or both odd, the only solution is $n=\nb$ and $m=\mb$.
Using
$$\int_{0}^{1}d\a \sum_{n\in Z} F(n+\a)=\int_{-\infty}^{\infty}d\a
F(\a)\eqno(3.24)$$
we finally obtain
$$\int_{0}^{1}d\a\int_{0}^{1}d\b Z_{\a}^{\b}(q,\qb;V)={\pi\over \sqrt{kIm
\tau}} \sum_{l,{\bar l}=0}^{k}\sum_{m=-k+1}^{k}N^{l,{\bar l}}c_{m}^{l}(q)
{\bar c}^{\bar l}_{m}(\qb)\eqno(3.25)$$
To obtain the full partition function for the coset we have to multiply
(3.25) with the contribution  from the local part of the gauge field
, $(det'${\square}$)^{-1/2}$ and the FP determinant, $(det'${\square}$)$ giving
a net contribution $(det'${\square}$)^{1/2}
=|\eta(q)|^{2}$ and an extra factor of $\sqrt{Im\tau}$ coming from the
measure of the twists (This factor is standard and can be read from the
norm of the gauge field
$|\delta A|^{2}=\int \sqrt{g}g^{\mu\nu}\delta A_{\mu}\delta A_{\nu}$
using the proper flat metric for a torus parametrized by $\tau$).
Putting everything together we obtain (up to constants)
$$Z^{V}_{SU(2)/U(1)}=|\eta(q)|^{2}\sum_{l,{\bar l}=0}^{k}\sum_{m=-k+1}^{k}
N^{l,{\bar l}}c_{m}^{l}(q){\bar c}_{m}^{\bar l}({\bar q})\eqno(3.26)$$
which is the correct parafermionic partition function, \cite{gq}.

Let us now consider the axial case.
The boundary condition (3.9) is replaced by
$$g(\sigma +1)=e^{i\pi\a\sigma_{3}}g(\sigma)e^{i\pi\a\sigma_{3}}\eqno(3.27)$$
{}From (3.10) we can verify that ${\bar J}^{\pm}$ is twisted with the
oposite sign of $\a$ compared to $J^{\pm}$.
Thus, the axial partition function is proportional to
$$\int_{0}^{1}d\a\int_{0}^{1}d\b Tr_{H}\left[q^{L_{0}(\a)}
\qb^{{\bar L}_{0}(-\a)}e^{2\pi i\b(J^{3}_{0}(\a)+{\bar J}_{0}^{3}
(-\a))}\right]\;.\eqno(3.28)$$
Doing the integrals over the moduli, we obtain in this case
$$Z^{A}={1\over 2}|\eta(q)|^2\sum^{k}_{l,{\bar l}=0}\sum_{m=-k+1}^{k}N^{l,{\bar
l}}c^{l}_{m}(q)({\bar c}^{\bar l}_{-m}({\bar q})+{\bar c}^{\bar
l}_{-m-2k}({\bar
q}))\eqno(3.29)$$

Using again the symmetry under the affine Weyl group (2.26) we obtain that
$$Z^{A}=Z^{V}\eqno(3.30)$$

One final comment is in order here, concerning the SU(2)/U(1) case:
We can also compute the parafermionic partition function $Z_{SU(2)/U(1)}(r,s)$
twisted around the two cycles of the torus by two elements of its parafermionic
symmetry $Z_{k}\times{\tilde Z}_{k}$, ($e^{2\pi i r/k}$, $e^{2\pi i s/k}$).
The way to do this is to allow a general twist
$$g(\sigma+1)=e^{i\pi\a\sigma_{3}}g(\sigma)e^{i\pi{\bar\a}\sigma_{3}}
\eqno(3.31)$$
with $k(\a-{\bar\a})/2=r$ mod $k$.
We must also project in the time direction on $J-{\bar J}=s$ mod $k$.
Since the twist is now neither axial nor vector there is the standard modular
anomaly that can be cancelled by multiplying the twisted partition function
by $exp(\pi k |\tau|^2(\a-{\bar\a})^2/4Im\tau)$.

The procedure described above for the $SU(2)/U(1)$ coset easily generalizes.
Consider a simple group $G$. We will gauge the maximal abelian subgroup,
namely the Cartan subalgebra. Thus we will be looking at the theory
of generalized parafermions, \cite{g}.
A convenient basis to work with is the Chevaley basis.
Let $J^{i}$ be a basis of the Cartan, and $\va,\va_{i}\in M$ denote the
roots and simple roots respectively.
The zero modes of the currents in this basis satisfy
$$[J^{i},J^{j}]=0\;\;,\;\;[J^{i},J^{\va}]={2\va\cdot\va_{i}\over
\va^{2}}J^{\va}
\;\;,\;\;[J^{\va},J^{-\va}]=\sum_{i}m_{i}J^{i}\eqno(3.32a)$$
$${2\va\over \va^{2}}=\sum_{i}m_{i}{2\va_{i}\over
\va_{i}^{2}}\;\;,\;\;[J^{\va},J^{\vb}]=\varepsilon_{\va,\vb}r_{\va,\vb}
J^{\va+\vb}\;\;{\em for}\;\va+\vb\in M\eqno(3.32b)$$
$$[J^{\va},J^{\vb}]=0\;\;{\em for}\;\;\va+\vb\;\notin M\eqno(3.32c)$$
where $r_{\va,\vb}$ is the smallest integer such that $\vb-r\va\notin M$ and
$\varepsilon_{\va,\vb}=\pm 1$.
The non-zero components of the Killing form in this basis are given by
$$\k(J^{i},J^{j})\equiv \k_{ij} ={4\va_{i}\cdot\va_{j}\over
\va_{i}^{2}\va_{j}^{2}}\;\;,\;\;\k(J^{\va},J^{-\va})={2\over
\va^{2}}\eqno(3.33)$$
Finally, the central term in the current algebra is given by $k$ times the
Killing form.

We can now impose the twisted boundary conditions similar to (3.9)
$$g(\s+1)=e^{2\pi iz_{i}T^{i}}g(\s)e^{-2\pi iz_{i}T^{i}}\eqno(3.34)$$
Their effect is to twist the algebra in the following way
$$J^{\va}_{m}\rightarrow J^{\va}_{m-s(\va)}\;\;,\;\;s(\va)=\sum_{i}z_{i}
{2\va\cdot\va_{i}\over \va^{2}}\eqno(3.35a)$$
$$J^{i}_{m}\rightarrow J^{i}_{m}+k(\sum_{j}\k_{ij}z_{j})\delta_{m,0}
\eqno(3.35b)$$
$$L_{m}\rightarrow L_{m}+\sum_{i}z_{i}J^{i}_{m}+{k\over 2}\left(\sum_{i,j}
\k_{ij}z_{i}z_{j}\right)\delta_{m,0}\;.\eqno(3.35c)$$
The projection factor now becomes
$$e^{2\pi i\sum_{i,j}w_{i}\k_{ij}(J^{j}_{0}-{\bar J}^{j}_{0})}\;.\eqno(3.36)$$
Using the string decomposition formulae, (2.24,25) (and properly accounting for
the change in the metric) we obtain

$$Z_{G}=\sum_{\Lambda,{\bar \Lambda}}\sum_{{\vec
\lambda}_{1,2}\in {M^{*}\over kM_{L}}}N^{\Lambda,{\bar \Lambda}}
c^{\Lambda}_{\vec\lambda_{1}}(q){\bar c}^{\bar \Lambda}_{\vec\lambda_{2}}
(\qb)\sum_{\va,\vb\in M_{L}}
q^{{k\over 2}(\va+{\vec z}+{\vec\lambda_{1}\over k})^{2}}\qb^{{k\over 2}
(\vb+{\vec z}+{\vec\lambda_{2}\over k})^{2}}e^{2\pi i{\vec w}\cdot
(k(\va-\vb)+{\vec\lambda}_{1}-{\vec\lambda}_{2})}.\eqno(3.37)$$
At this stage, inspection of (3.37) reveals that all we have found in the
SU(2) case goes through here.
In particular, a Weyl reflection generated by the simple root $\va_{i}$
is generating from (3.37) the partition function with the $U(1)$ subgroup
in that direction axially gauged.

An interesting point is that the (axial-vector) duality transformation
in the abelian coset can be effected also via an orbifold construction
,  (this has been observed for $SU(2)_{k}/U(1)$ in \cite{gq}).
Let us consider the parafermionic partition function on the torus with
boundary conditions around the two cycles twisted by elements
$e^{2\pi ir/k}$, $e^{2\pi is/k}$ of the $Z_{k}$ parafermionic symmetry.
This can be evaluated to be
$$Z(r,s)={1\over 2}|\eta|^{2}e^{-2\pi i{s\over
k}}\sum_{l=0}^{k}\sum_{m=-k+1}^{k}c^{l}_{m}{\bar c}^{l}_{m-2r}\eqno(3.38)$$
and the usual vector partition function is $Z(0,0)$.
If we construct the orbifold of the original theory with repect
to the $Z_{k}$ symmetry, (which amounts to summing over $r,s$),
we obtain the axial partition function.
In this respect the orbifold projection throws out the order operators
and adds as twisted sectors the disorder operators.
This is precisely the generalization of what is known to happen in
the Ising model.
This automatically generalizes to arbitrary abelian cosets where the
parafermionic symmetry group is isomorphic to $M^{*}/kM_{L}$.
Thus we have the following sequence. We gauge the vector U(1), and thus obtain
a model with a Killing symmetry associated with axial U(1). This U(1) symmetry
is broken to a discrete group (the parafermionic symmetry).
Doing an orbifold on that symmetry (which amounts to a flat gauging) we obtain
the axially gauged theory.

\section{$O(d,d)$ symmetries}

Combining duality transformations with antisymmetric tensor shifts
and linear transformations on the Cartan angles, we can generate
a bigger "duality" group which at the level of the partition function
acts as $O(d,d,Z)$, \cite{gr}.
In the case of toroidal backgrounds it is easy to see how this works.
Invariance under $B_{ij}\rightarrow B_{ij}+N_{ij}$ is obvious from
(1.14c), where $N_{ij}$ is an antisymmetric matrix with integer
entries. Also, invariance under $G+B\rightarrow U(G+B)U^{T}$ is obvious
from (1.10), where $U$ is an arbitrary matrix with integer entries.
The duality group (1.9) and the transformations above generate the $O(d,d,Z)$
group.

In the non-flat case, similar arguments apply, \cite{gr}, with one difference:
the duality group acting on the full operator content of the theory is more
complicated than its reduction on the partition function. This is the
reflection of our observation that the full duality group in the non-abelian
case is isomorphic to the finite Weyl group (or its reductions by subgroups).
Keeping this in mind, we can reproduce easily the argument for the partition
function.
The most general $\s$-model action with $d$ chiral currents is of the form
(2.10). We will rewrite it in chiral form, and we will explicitly
parametrize the $I(h)$. We will be a bit more general than \cite{gr}
by allowing arbitrary radii for the Cartan angles.
The general action takes then the form (up to total derivatives)
$$S={1\over 4\pi}\int \left[\kappa_{ij}(\p\a^{i}\pb\a^{j}+\p\g^{i}\pb\g^{j})+
2\Sigma_{ij}(x)\p\a^{i}\pb\g^{j}+\Gamma^{1}_{ai}(x)\p x^{a}\pb\g^{i}+
\Gamma^{2}_{ia}(x)\pb x^{a}\p\a^{i}+\right.$$
$$\left. +\Gamma_{ab}(x)\p x^{a}\pb x^{b}
\right]-{1\over 8\pi}\int R^{(2)}\Phi(x)\eqno(4.1)$$
where $\a^{i},\g^{i}$ take values in $[0,2\pi]$.
The action (4.1) is invariant under $\a^{i}\rightarrow \a^{i}+\varepsilon^{i}(
{\bar z})$
and $\g^{i}\rightarrow \g^{i}+\zeta^{i}(z)$ with associated chiral (abelian)
currents,
$$J^{i}=\p\g^{j}\kappa_{ji}+\p\a^{j}\Sigma_{ji}+{1\over 2}\p
x^{a}\Gamma^{1}_{ai}\eqno(4.2a)$$
$${\bar J}^{i}=\kappa_{ij}\pb\a^{j}+\Sigma_{ij}\pb\g^{j}+{1\over
2}\Gamma^{2}_{ia}\pb x^{a}\;.\eqno(4.2b)$$
This automatically implies (assuming conformal invariance) that $S$ describes
a (not direct, in general) tensor product of a WZW model and some arbitrary
decoupled CFT.
The currents (4.2) generate the Cartan subalgebra of the full current algebra
of the WZW model.
We can gauge vectorially the Cartan subalgebra,
$$S_{V}=S+{1\over 4\pi}\int\left[A^{i}{\bar J}^{i}-{\bar A}^{i}J^{i}+
{1\over 2}A^{i}(\kappa-\Sigma)_{ij}{\bar A}^{j}\right]\;.\eqno(4.3)$$
Integrating out the gauge fields and gauge fixing $\a^{i}=\g^{i}$ we obtain
$$S_{c}={1\over 4\pi}\int\left[E_{ij}(x)\p\a^{i}\pb\a^{j}+F^{1}_{ai}(x)\p x^{a}
\pb\a^{i}+F^{2}_{ia}(x)\p\a^{i}\pb x^{a}+F_{ab}(x)\p x^{a}\pb
x^{b}\right]-{1\over
8\pi}\int R^{(2)}\phi(x)\eqno(4.4)$$
where
$$E_{ij}(x)=4\kappa
(1+\kappa^{-1}\Sigma)(1-\kappa^{-1}\Sigma)^{-1}\eqno(4.5a)$$
$$F^{2}(x)=2\kappa(\kappa-\Sigma)^{-1}\Gamma^{2}\;\;,\;\;F^{1}(x)=2\Gamma^{1}
(\kappa-\Sigma)^{-1}\kappa\;\;,\;\;F=\Gamma-{1\over 2}\Gamma^{1}
(\kappa-\Sigma)^{-1}\Gamma^{2}\eqno(4.5b)$$
$$\phi(x)=\Phi+{\rm log}({\rm det}(\kappa-\Sigma))\;.\eqno(4.5c)$$

The interesting observation is that, given a $\s$-model (4.4) with $d$
Killing symmetries we can always construct it as an abelian coset of a WZW
model (4.1).
The reason is that relations (4.5) are generically invertible. There is an
underlying assumption in this, that should be kept in mind, namely that
conformal
invariance of (4.4) implies conformal invariance of (4.1).

The duality generators $D_{i}$ correspond to switching from vector to axial
in the $i$-th component of the gauging.
There are also the following obvious symmetries, integer shifts of the
antisymmetric tensor $E_{ij}(x)\rightarrow E_{ij}(x)+N_{ij}$ with $N$ an
antisymmetric
integer matrix, and integer linear transformations of the angles $\a^{i}$
which act as $E\rightarrow UEU^{T}$, $F^{2}\rightarrow UF^{2}$,
$F^{1}\rightarrow F^{1}U^{T}$.
The full group of invariance of the partition function is the $O(d,d,Z)$
group acting as
$$\left(\matrix{E& F^{2}\cr F^{1}&F\cr}\right)\rightarrow
\left(\matrix{(aE+b)(cE+d)^{-1}&(ac^{-1}d-b)(cE+d)^{-1}cF^{2}\cr
F^{1}(cE+d)^{-1}&F-F^{1}(cE+d)^{-1}cF^{2}\cr}\right)\eqno(4.6)$$
where
$$\left(\matrix{a&b\cr c&d\cr}\right)\;\in\; O(d,d,Z)\eqno(4.7)$$

The exact underlying picture of the symmetries above is as follows.
The antisymmetric tensor shift in the action corresponds to combined affine
Weyl translations on the left and right parts of the theory.
The duality transformations, as we argued in the previous section are
isomorphic to Weyl transformations.
Finally the $GL(d)$ group acts a linear integer transformations of the weight
lattice.
Again, these are exact symmetries at least when the coset is compact.

The reasoning above can be extended to derive the $O(d,d,R)$ action on
conformal backgrounds.
This was first observed as an invariance of the one-loop string effective
action, with backgrounds having $d$ Killing symmetries, \cite{ven}.
The observation is the following. Starting from a background (CFT) with
$d$ Killing symmetries, an arbitrary
constant shift of the antisymetric tensor, as well as an arbitrary linear
combination of the coordinates corresponding to the Killing directions
provide another theory which is also conformally invariant.
If these transformations are intertwined with the duality transformations
$D_{i}$ it can be shown that the full group of transformations is isomorphic
to $O(d,d,R)$ which acts as in (4.6).
The $O(d,d,Z)$ subgroup generates the same string theory.
It is obvious that linear transformations and antisymmetric tensor shifts
are exact to all orders in $a'$ and the string loop expansion.
In the compact case we have shown in the previous sections that, although the
action of the duality
transformations $D_{i}$ has to be modified beyond one-loop, there is
such a modification, that is exact again non-perturbatively in $\a '$
and perturbatively in the string loop expansion.\footnote{Arguments,
to the extend that $O(d,d,R)$ transformations can be made exact symmetries
to all orders in $\a'$ where also given in \cite{sen} from a string field
theory point of view.}

One further comment applicable to the compact case: $O(d,d,R)$ transformations
do not in general preserve the positivity (unitarity in Minkowski space) of the
appropriate conformal field theory. This is obvious for antisymmetric tensor
shifts, since they correspond to arbitrary shifts of the weight lattice and
thus map integrable to non-integrable reps. It remains to be seen if the
string theory constructed from such CFTs remains unitary.

\section{On marginal current-current perturbations}

In a CFT with chiral abelian currents like (4.1) it is well known \cite{cs}
that the perturbation
$$S_{I}=\lambda\int g_{ij}J^{i}{\bar J}^{j}\eqno(5.1)$$
is marginal.
This corresponds to the deformation of the Cartan torus, and generalizes
the SU(2) case which we explicitly discussed in section 2.
It is also well known from CFT that such perturbations break the non-abelian
symmetry while leaving the (abelian) Cartan symmetries intact.
However a simple calculation shows that the action $S+S_{I}$ has (deformed)
chiral symmetries only to order $\cal{O}(\lambda)$.
The way to improve this situation is via (in this case) a special $O(2d,2d)$
transformation, or equivalently by considering the tensor product of this
theory with $d$ free scalar fields and gauging an arbitrary linear combination
of the two $U(1)^{d}$ symmetries.\footnote{Similar observations were made
independently in \cite{hs} and \cite{nh}.}

To see that we can get the current-current perturbation from an $O(2d,2d)$
transformation we can study first infinitesimal perturbations.
Let us consider the infinitesimal form of the transformations in (4.6).
$$a\sim 1+\lambda A+{\cal{O}}(\lambda^{2})\;\;,\;\;d\sim 1-\lambda A^{T}+
{\cal{O}}(\lambda^{2})\eqno(5.2a)$$
$$b\sim \lambda B+{\cal{O}}(\lambda^{2})\;\;,\;\; c\sim \lambda
C+{\cal{O}}(\lambda
^{2})\;\;,\;\;B^{T}=-B\;\;,\;\;C^{T}=-C\eqno(5.2b)$$
It is not difficult to see that the following infinitesimal $O(2d,2d)$
transformation
$$\left(\matrix{1&\lambda g^{T}/4&0&-\lambda g^{T}/4\cr \lambda g/4&1&
\lambda g &0\cr
0&\lambda g^{T}&1&-\lambda g^{T}/4\cr -\lambda g/4&0&-\lambda g/4&1\cr}\right)
+{\cal{O}}(\lambda^{2})
\eqno(5.3)$$
generates the perturbation (5.1)
Of course there will be also a non-trivial dilaton that can be calculated
from (4.5c), which will ensure conformal invariance at the one-loop level.

For the $SU(2)$ case there is only one marginal perturbation and (5.3)
can be integrated automatically to obtain the finite transformation.
The advantage of the finite transformation is that the theory with the fully
transformed action will have chiral abelian currents for all values of the
parameters. However conformal invariance will still have to corrected
beyong one loop.

What we remarked so far  is hardly surprising.
When we look however at the action of $O(d,d)$ transformations on the abelian
coset theory, which generically has no chiral currents, we can observe
that it still implies that certain current-current perturbations are marginal.

Let us first compute the conserved currents in (4,4) associated to the Killing
symmetries $\a^{i}\rightarrow \a^{i}+\varepsilon^{i}$
$$ J^{i}=\p\a^{j}E_{ji}(x)+\p x^{a}F^{1}_{ai}(x)\eqno(5.4a)$$
$$ {\bar J}^{i}=E_{ij}(x)\pb\a^{j}+F^{2}_{ia}(x)\pb x^{a}\;.\eqno(5.4b)$$
These currents are conserved
$$\pb J^{i}+\p{\bar J}^{i}=0\eqno(5.5)$$
but not chirally conserved.

The infinitesimal transformations corresponding to (4.6) become
$$\delta E=\lambda (AE+EA^{T}+B-ECE)\;\;,\;\;\delta F=-\lambda
F^{1}CF^{2}\eqno(5.6a)$$
$$\delta F^{1}=\lambda F^{1}(A^{T}-CE)\;\;,\;\;\delta F^{2}=\lambda (A-EC)F^{2}
\;.\eqno(5.6b)$$

We can now observe that the infinitesimal change in the action (4.4) under
the special transformation $A=B=0$ has the form
$$\delta S =-\lambda \int J^{i}C_{ij}{\bar J}^{j}\eqno(5.7)$$
which is a specific current-current perturbation (because the matrix $C$ is
forced by (5.2b) to be antisymmetric).
Of course, there are corrections again to the dilaton via (4.5c).

In CFT, whenever there are conserved but not chirally conserved currents,
they are bad conformal fields. This can be proven in general in 2-d
by showing that normal conservation of a current and conformal invariance
(which fixes the form of the two-point function) implies chiral conservation.
Moving a bit off criticality we can see that conserved but not chirally
conserved
currents have severe IR divergences and decouple from the spectrum as one
approaches the critical point.
It is thus surprising that a perturbation of the form (5.7) is a marginal
perturbation of such models.

\section{Non-compact cosets}

Non compact cosets attracted attention recently, \cite{wit2,giv,dvv,k}
as CFTs that provide curved backgrounds for consistent string propagation.
They also generically exhibit (semi-classically) spacetime singularities.

The prototype theory (and apparently the simplest) is the $SL(2,R)/U(1)$
model describing a two dimensional target manifold.
We will consider Euclidean targets, which means that the U(1) we are going
to gauge will be compact.
If we parametrize the $SL(2,R)$ matrix using Euler angles as
$$g=e^{i{\phi\over 2}\s_{2}}e^{{r\over 2}\s_{1}}e^{i{\psi\over
2}\s_{2}}\eqno(6.1)$$
then, upon integrating out the gauge fields, we arrive at the following
partition functions for the axial and vector theory
$$Z_{A}=\int_{0}^{\infty}{{\rm sinh}rdr\over 1+{\rm cosh}r}\int_{0}^{2\pi}d\phi
e^{-{k\over 4\pi}\int[\p_{\mu}r\p^{\mu}r+4{\rm tanh}^{2}{r\over 2}\p_{\mu}\phi
\p^{\mu}\phi]}\eqno(6.2)$$
$$Z_{V}=\int_{0}^{\infty}{{\rm sinh}rdr\over 1-{\rm cosh}r}\int_{0}^{2\pi}d\phi
e^{-{k\over 4\pi}\int[\p_{\mu}r\p^{\mu}r+4{\rm coth}^{2}{r\over 2}\p_{\mu}\phi
\p^{\mu}\phi]}\eqno(6.3)$$
where we have incorporated the dilaton into the measure.
In the axial theory, the string propagates on a manifold with the shape of a
cigar, which becomes a cylinder asymptotically ($r\rightarrow\infty$).
However the manifold of the vector theory, although similar when $r\rightarrow
\infty$, has a different, and in fact singular behaviour as $r\rightarrow 0$.
The line element and scalar curvature behave as follows, in this region
$$ds^{2}\sim dr^{2}+{1\over r^{2}}d\phi^{2}\;\;\;;\;\;\;R\sim {1\over r^{2}}
\;.\eqno(6.4)$$
In the Minkowskian (2-d black hole) model the analytic continuation of the
axial Euclidean model (6.2) generates region I of spacetime (the asymptotically
flat region till the horizon). Region III (from the horizon to the sigularity)
corresponds to the self-dual SU(2)/U(1) model. Finally region V (behide the
singularity) corresponds to the vector Euclidean model (6.3).

The pertinent question here is: are the two models (6.2,3) equivalent,
like in the compact case? Our semiclassical derivation of the axial to vector
duality is still valid here. However there are reasons to make us distrustful
of such a semiclassical reasoning in the non-compact case.
One is that the two targets are radically different , unlike the compact case
where the target manifold of the axial theory  is a reparametrization
of that of the vector theory (even when higher loop corrections are included,
\cite{dvv}).
The other reason is that the semiclassical spectrum of the two theories
in the non-compact case is quite different.
In the axial theory, only the continuous series of the SL(2,R) representations
contribute, while in the vector theory there are extra contributions from
the discrete series.
The partition function of the axial model has been computed by Gawedski
\cite{gaw}, however that of the vector model remains a mystery.

Trying to understand the situation, we will analyse a simpler (but not trivial)
case of (potential) axial-vector duality in a non-compact model.
Let us consider the conformal field theory on a 2-d Euclidean plane
$$Z_{E}=\int d^{2}x e^{-{1\over 4\pi}\int[(\p x^{1})^{2}+(\p
x^{2})^{2}]}\eqno(6.5)$$
This is a free field theory that we know everything about, in particular,
its exact torus partition function is (up to constants)
$$Z^{t}_{E}={1\over Im\tau}{1\over |\eta(\tau)|^{4}}\eqno(6.6)$$
where the $1/Im\tau$ factor comes from the integration of the zero modes.
We can write this theory in polar coordinates $(r,\t)$,
$$Z_{E}=\int_{0}^{\infty}rdr\int_{0}^{2\pi}d\t \; e^{-{1\over 4\pi}\int[(\p
r)^{2}+
r^{2}(\p \t)^{2}]}.
\eqno(6.7)$$
We can now apply the O(1,1) duality transformation corresponding to the Killing
symmetry associated with translations of $\t$ to obtain the ``dual" theory
$$Z_{\e}=\int_{0}^{\infty}{dr\over r}\int_{0}^{2\pi}d\t \; e^{-{1\over
4\pi}\int[
(\p r)^{2}+r^{-2}(\p \t)^{2}]}\eqno(6.8)$$
where the effects of the dilaton transformation were also taken into account
by the change in the measure. This is enough for the bulk effects. When we
consider
manifolds with boundary there are extra corrections due to the coupling of the
dilaton
term to the extrinsic curvature of the boundary. Such coupling will be
important
when we later consider the calculation of the propagator on the 2-punctured
sphere,
or equivalently the cylinder.
A naive extrapolation of our results from the compact case would imply that
these theories are equivalent, and in particular that the theory (6.8) is a
free
field theory.
However, as in the SL(2,R)/U(1) example, the manifold corresponding to (6.8)
is radically different from the flat Euclidean plane of (6.6); it is a curved
manifold with a curvature singularity at the origin. It concides with a region
close to the origin, of the vector SL(2,R)/U(1) model as can be seen from
(6.4).

It is also interesting to note that the two theories (6.6,8) can be viewed
as axial and vector gauged models of the following $\s$-model, with 3-d target
$$Z_{3}=\int_{0}^{\infty}g(r)dr\int_{0}^{2\pi}d\t d\varphi \;\; e^{-{1\over
4\pi}\int[(\p \t)^{2}+(\p \varphi)^{2}+(\p
r)^{2}+2f(r)\p\t\pb\varphi]}\eqno(6.9)$$
Choosing
$$g(r)={r\over 1+r^{2}}\;\;\;,\;\;\;f(r)={1-r^{2}\over 1+r^{2}}\eqno(6.10)$$
we can verify with a simple computation that by gauging the axial symmetry
$\t\rightarrow\t+\varepsilon$, $\varphi\rightarrow\varphi+\varepsilon$ we
obtain the free model (6.7) while gauging the vector symmetry
$\t\rightarrow\t+\varepsilon$, $\varphi\rightarrow\varphi-\varepsilon$ we
obtain model (6.8).
It is an interesting question whether the the model (6.9,10) is conformally
invariant.
There are abelian chiral currents in this model
associated with the symmetries $\t\rightarrow\t+\varepsilon({\bar z})$ and
$\varphi \rightarrow \varphi+\zeta(z)$
$$J=\p\varphi+f(r)\p\t\;\;,\;\;{\bar J}=\pb\t+f(r)\pb\varphi\;\;;\;\;\p{\bar
J}=
\pb J=0\eqno(6.11)$$
We can easily check that unless $f(r)$ is the one which corresponds to the
SL(2,R) model there are no other chiral currents in the theory (this might
seem trivial, but it is possible in principle that the model can be mapped
to that of SL(2,R) through a complicated reparametrization).
Thus if (6.9) is conformally invariant it describes the product (certainely
not direct) of a U(1) theory and some other CFT.

We will now proceed to tackle the question posed above: is the free model
(6.6) and (6.8) equivalent?
We can proceed by first studying the symmetries of the two models.
The flat model posseses $U(1)\times U(1)$ current algebra. The two left U(1)
currents are given by $\p x^{1}$ and $\p x^{2}$ in the Cartesian basis.
The dual model (6.8) has a symmetry that comes quite close to current
algebra: ``parafermionic symmetry".
One can easily verify, using the equations of motion, that the (non-local)
parafermion currents
$$\psi_{\pm}=\p_{z}\left[ r\exp \left[\pm i\int {\pb \t\over r^2}d{\bar z}-
{\p \t\over r^2}dz\right]\right]\eqno(6.12a)$$
$${\bar \psi_{\pm}}=\p_{\bar z}\left[ r\exp \left[\pm i\int {\pb \t\over
r^2}d{\bar z}-
{\p \t\over r^2}dz\right]\right]\eqno(6.12b)$$
are chirally conserved
$$\p {\bar \psi_{\pm}}=\pb \psi_{\pm}=0\eqno(6.13)$$
As with the standard $\s$-model picture of parafermions, they do not
depend on the position of the string, due to the $U(1)$ symmetry
associated with translations in $\t$.
The parafermions (6.12) coincide with the $k\rightarrow\infty$ limit
of $SL(2,R)_{k}/U(1)$ parafermions \cite{bbkk} and thus they should behave as
chiral currents (that is their exact conformal weight is 1).
An map between the symmetries of the two models provides a map between their
classical solutions.
This map exists both for Minkowski and Euclidean signature of the target space.
Given any solution of the equations of motion of the flat model, $\p\pb
x^{1}=0$,
$\p\pb x^{2}=0$, we can construct a unique classical solution of the dual model
as follows
$$r^2=(x^{1})^{2}+(x^{2})^{2}\eqno(6.14a)$$
$$\p\t =(x^{2}\p x^{1}-x^{1}\p x^{2})\;\;\;,\;\;\;\pb\t=-(x^{2}\pb x^{1}-x^{1}
\pb x^{2})\eqno(6.14b)$$
Notice that the system (6.14b) is compatible (integrable).
It is highly possible that this map generates all the solutions of the dual
model.

In the Minkowsi case the map becomes :
$$r=(x^{1})^{2}-(x^2)^{2}\eqno(6.15a)$$
$$\p\t =(x^{2}\p x^{1}-x^{1}\p x^{2})\;\;\;,\;\;\;\pb\t=
(x^{2}\pb x^{1}-x^{1} \pb x^{2})\eqno(6.15b)$$
In this case $r$ has indefinite sign.

The Minkowski version of the dual model gives a very interesting spacetime
picture.
The metric in lightcone coordinates is $ds^{2}={dudv\over u+v}$
whereas the scalar curvature is $R\sim {1\over u+v}$.
There is a curvature singularity at $u+v=0$. One can solve for the
geodesics on this manifold
$$u(t)=A^{2}e^{\Gamma t}+\Gamma ABt+AB+\Delta\eqno(6.16a)$$
$$v(t)=B^{2}e^{-\Gamma t}-\Gamma ABt+AB-\Delta\eqno(6.16b)$$
where $A,B,\Gamma,\Delta$ are arbitrary complex numbers subject only
to the constaint that $u(t),v(t)$ should be real for all $t$.
One can see that the geodesics that reach the singularity are reflected by it.
Depending on the choice of what to call time, this metric describes either
an eternal singularity at some space point (black hole picture) or a
singularityover all space at some value of time (``big-bang" picture).
We should also remark that there is no horizon here.
The two regions on opposite sides of the singularity correspond to the
$k\rightarrow \infty$ limit (in the $SL(2,R)_{k}/O(1,1)$ model ) of the region
between the horizon and the singularity, as well as the region
behide the singularity. The flat model is the limit
of the asymptotic region outside the horizon.

Let us return to the question of equivalence of the two theories beyond
the classical level. The first object to compare would be the partition
functions.
Unfortunately it seems extremely difficult to compute exactly the torus
partition function of the $\e$ model. Thus we will resort to an improved
version of the so called
``minisuperspace" approximation. This approximation amounts essentially to a
dimensional
reduction to 1-d, that is, neglecting the $\s$ dependence. Thus, we will have
to
deal
with a quantum mechanical model with Langrangian given by
$$L_{\e}={1\over 4\pi}[{\dot r}^{2}+r^{-2}{\dot \t}^{2}]\eqno(6.12)$$

The minisuperspace approximation of a CFT is not an approximation in the usual
sense of the word.
However, it is well understood that, since it describes the quantum mechanics
of zero modes, it does not ``see" the oscilator part of the spectrum,
finite renormalizations of couplings and unitary truncations of the Hilbert
space as well as toplogically different sectors.
For example in the ``minisuperspace" approximation to the $SU(2)_{k}$ WZW
model all representations of $SU(2)$ contribute whereas in the 2-d theory
their range is restricted to $0\leq j\leq k/2$.
Here we will need an improved version of this though since even in case
of the $R\rightarrow 1/R$ duality, naive dimensional reduction misses
the winding states.
There one has in fact to perform a 1-d path integral and sum over backgrounds
with linear $\s$ dependence. In this approximation duality is manifest.

Thus we will consider $\t =m\s +\t (\tau )$, $r=r(\tau)$. Since $\t$ is a
function
on the cylinder, $m\in Z$.
Then,
$$Z_{E}=\sum_{m\in Z}\int e^{-S_{m}}\;\;,\;\;Z_{\tilde E}=\sum_{m\in Z}\int
e^{-{\tilde S}_{m}}
\eqno(6.18)$$
with
$$S_{m}=\int d\tau [{\dot r}^{2}+r^{2}{\dot
\th}^{2}+m^{2}r^{2}]\;\;,\;\;{\tilde S}_{m}=
\int d\tau [{\dot r}^2 +{1\over r^2}{\dot \th}^{2} +{m^{2}\over
r^2}]\eqno(6.19)$$

In order to compare the two systems we will derive the Hamiltonians responsible
for the path integrals above. The method is standard and relies in deriving a
partial
differential equation for quantum mechanical amplitute $F(x,x',T)$ for a
particle to go from $x$
to $x'$ at time $T$.
The quantum mechanical partition function can be obtained from $F(x,x',T)$ by
taking
the trace, that is setting $x=x'$ and integrating over all space.
The amplitude though contains more information than the partition function.
We use the conventional measure in the path integral, $\sqrt{G}$, where
$G_{\mu\nu}$ is
the target space metric  but we will include also a factor
$[\sqrt{G(t_{init})}/\sqrt{G(t_{final})}]^{s}$ with $s=1/4$ which comes from
the coupling of the dilaton to the boundaries of the 2-d cylinder (this exists
only for the dual model).
Then a standard computation (see for example \cite{hh}) shows that the
amplitude satisfies
the following partial differential equation

$${\p \over \p T}F=\left[ G^{\mu\nu}\nabla_{\mu}\nabla_{\nu}+{1\over
3}(s-1)R-V\right]F\eqno(6.20)$$
where $V$ is the potential.
Thus, the right hand side of (6.20) is identified with minus the Hamiltonian.

For the flat model the curvature is zero and there is no dilaton thus the
Hamiltonian is
$$H_{E}=-\left[ {\p^{2}\over \p r^{2}}+{1\over r}{\p \over \p r}+{1\over
r^{2}}{\p^{2}\over
\p\t^{2}}-m^{2}r^{2}\right]\eqno(6.21)$$
The measure for the wavefunctions is $\int_{0}^{\infty}rdr\int_{0}^{2\pi}$.
For a given angular momentum sector, $\psi(r,\th)=R(r)e^{in\th}$, the effective
Hamiltonian
on $R(r)$ is
$$H^{eff}_{E}=-\left[ {\p^{2}\over \p r^{2}}+{1\over r}{\p \over \p
r}-{n^2\over r^{2}}
-m^{2}r^{2}\right]\eqno(6.22)$$

For the dual model the scalar curvature is $R=-4/r^{2}$ and the boundary
effects of the dilaton
give $s=1/4$, so
$$H_{\tilde E}=-\left[ {\p^2\over \p r^2}-{1\over r}{\p \over \p
r}+r^2{\p^2\over \p \th^{2}}+
{(1-m^2)\over r^2}\right]\eqno(6.23)$$
and the measure on the wavefunctions is $\int_{0}^{\infty}dr/r\int_{0}^{2\pi}$.
Again for a given angular momentum sector, ${\tilde \psi}={\tilde
R}(r)e^{in\th}$, the effective Hamiltonian becomes
$$H^{eff}_{\tilde E}=-\left[ {\p^2\over \p r^2}-{1\over r}{\p \over \p
r}+-n^{2}r^2+
{(1-m^2)\over r^2}\right]\eqno(6.24)$$
If we now redefine $S(r)=\sqrt{r}R(r)$ and ${\tilde S}(r)={\tilde
R}(r)/\sqrt{r}$ so that
the measures for both $S,{\tilde S}$ become flat, $\int_{0}^{\infty}dr$, then
the two effective Hamiltonians become
$${\hat H}_{E}^{eff}=-\left[ {\p^{2}\over \p r^{2}}+{{1\over 4}-n^2\over r^{2}}
-m^{2}r^{2}\right]\eqno(6.25a)$$
$${\hat H}^{eff}_{\tilde E}= -\left[{\p^{2}\over \p r^{2}}+{{1\over 4}-m^2\over
r^{2}}
-n^{2}r^{2}\right]\eqno(6.25b)$$

It is obvious that the two models have identical spectra upon the interchange
of $m\leftrightarrow n$.
This not only shows that their partitions functions are equal (in this
approximation) bu provides also the exact map between the states.
A comparison with the exact CFT propagator in the flat model, \cite{pol},
indicates that the
point-like models (zero modes) of the dual model correspond to certain
oscilator
modes in the flat model.
This confirms that a symmetry similar to affine Weyl translations is also at
work here.
Thus so far we have verified that duality is a symmetry in two approximations:
the semiclassical \cite{k} and the improved minisuperspace approximation.
Since the nature of the two approximations is different, it is highly plausible
that duality.
even in the non-compact case, is an exact symmetry.
It turns out that there is a way of proving axial-vector duality for
non-compact cosets.
This will be presented elsewhere, \cite{gk}.

A similar analysis can be performed in the more ``realistic" SL(2,R)/U(1)
model, (6.2,3), and
similar remarks apply there.
As we mentioned earlier the two versions (6.2) and (6.3) differ substantially
only in a neighbourhood of $r=0$, and there they are approximated
by our toy models (6.7) and (6.8).

The indications from our analysis imply that affine Weyl symmetry is also
at work in the non-compact case.
This is however a highly non-trivial statement at the level of affine
representations.
It is not difficult
to see that in the non-compact case, affine Weyl translations map in general
a representation to a different one.
This can be seen at the level of the non-compact string functions, \cite{bk}.
However, duality idicates that we should consider orbits under the translation
group, but in
that  case one has always to cope with non-positive representations and the
spacetime interpretation is not manifest.
One way of this might be the suggestion in \cite{vafa}.

In many issues associated with black-holes one usually invokes some analytic
continuation from Minkowski to Euclidean space. As we have seen it plausible
that there are two inequivalent such continuations depending on the
region of spacetime. This might imply a different behaviour (and maybe
interpretation) for such issues as Hawking radiation etc.

\centerline{NOTE ADDED}

After the completion of this work I have found that the connection between
the duality transformations in k=1 (toroidal) WZW models
for simply laced groups and the finite Weyl group (a special case of what
has been discribed here) has been noticed before, \cite{can}.
I would like to thank A. Schwimmer for bringing this to my attention.

\end{document}